\documentstyle[preprint,aps,prd]{revtex}
\newcommand{\beq}{\begin{equation}}
\newcommand{\eeq}{\end{equation}}
\newcommand{\bea}{\begin{eqnarray}}
\newcommand{\eea}{\end{eqnarray}}
\newcommand{\fm}{\tilde F_{\mu}}

\newcommand{\fr}{\tilde F_{\rho}}
\newcommand{\fmn}{F_{\mu\nu}}
\newcommand{\oh}{\displaystyle{\frac{1}{2}}}
\begin{document}
\title          {Effective Theory for
               Parity Conserving $QED_3$}
\author{
I.J.R. Aitchison\thanks{ijra@thphys.ox.ac.uk}}

\address{ Department of Physics, Theoretical Physics,
University of Oxford\\
1 Keble Road, Oxford OX1 3NP, UK}

\author{C.D. Fosco\thanks{fosco@cab.cnea.edu.ar}}

\address{ Centro At\'omico Bariloche\\
8400 Bariloche, Argentina}

\author{F.D. Mazzitelli\thanks{fmazzi@df.uba.ar}}

\address{ Departamento de F\'\i sica and IAFE\\
 Facultad de Ciencias Exactas y Naturales\\
 Ciudad Universitaria, Pabell\' on 1\\
 1428 Buenos Aires, Argentina}

\maketitle
\begin{abstract}

We consider a higher derivative effective theory for an Abelian gauge
field in three dimensions, which represents the result of integrating
out heavy matter fields interacting with a
classical gauge field in a parity-conserving way.  We retain terms
containing up  to
two derivatives of $F_{\mu\nu}$, but make no assumption about  the
strength of this field. We then quantize the gauge field, and compute
the one-loop  effective action for a constant $\fmn$. The result
is explicitly  evaluated for  the case  of a constant
magnetic field.
\end{abstract}

\bigskip

\section{Introduction}

$QED_3$-like theories, namely those consisting of a charged matter field
in interaction with  an Abelian gauge  field in $2+1$  dimensions, have been
the subject of intense research  in recent years, because of  their multiple
applications to both Condensed  Matter and High Energy  Physics~\cite{frad}.
The specific form of the action chosen for matter and gauge fields  depends,
of course, on the particular  system one wants to represent.  Although gauge
invariance is almost exclusively assumed, both  parity-violating~\cite{pavi}
and parity-conserving~\cite{paco} actions have been considered, depending on
the  realm  of  application.  Given  the  large  number  of   models one can
construct within each  of these categories,  the task of  identifying their
common dynamical properties  would be quite  difficult if accomplished  on a
case by case basis.

The aim of this  paper is to construct  a unified description for  a general
gauge-invariant and parity-conserving  $QED_3$-like model, assuming  massive
matter  fields.
The combination of  all the different models under a  common
description  is  made  possible  by  using  the  `coarse  graining' provided
by a low-momentum approximation.
Our approach consists of  constructing a `classical' effective
theory  for the gauge  field $A_\mu$, which can be interpreted as arising
after integrating out the massive matter fields~\cite{cang}, keeping
a finite number  of derivatives acting on $F_{\mu\nu}$.
We may call the  theory for $A_\mu$ at  this
stage `classical', since $A_\mu$ is not yet quantized. The next step amounts
to  adding  to  this  `classical'  effective  theory  the  quantum  effects due
to  virtual  photons~\cite{halt},  calculating   the  corresponding  one-loop
`quantum effective action'. We shall compute this effective action for a
constant (but arbitrarily strong) $F_{\mu \nu}$, and then evaluate it
explicitly for the case of a constant magnetic field or arbitrary intensity.

The (at first  sight) dangerous features  introduced by the  higher derivative
character of the classical effective theory are
dealt with by consistently cutting off
the Euclidean momenta in the gauge field loops at values of the order of the
mass of the matter fields, since the higher derivative theory is reliable in
that  low-momentum  region  only.  This  procedure  avoids  integrating over
regions where unphysical poles in the gauge field propagator could show  up.
These unphysical poles, on the other hand, are artifacts of the low-momentum
approximation. We will show that,  had the  low-momentum approximation  not
been
made, the loop integrals would have been regular everywhere.

The procedure of starting from a general effective theory,
involving a small number of
arbitrary parameters, has the welcome feature that it allows us to keep  the
discussion as general  as possible, describing  many cases at  the same time
(namely, fermions and/or bosons, many fields, etc), since the result can  be
stated in terms of general parameters, the values of which in any particular
model can  be found  quite easily.

This paper  is organized  as follows:  In section  2 the classical effective
theory for the gauge field is constructed, while in section 3 the one-loop quantum
effective action for the gauge field is defined and
then calculated for a general
constant $\fmn$, obtaining explicit results for some simplified  situations.
In section 4  we discuss our  results and show  how to avoid  the use of the
cutoff by improving upon the use of the low-momentum approximation.

\section{The Model}

As partially advanced in the previous section, our three-dimensional
classical effective
action for the Abelian gauge field $A_\mu$ is determined by requiring
it to be a general gauge-invariant and parity-conserving one. Locality is
also assumed,
and this is justified by assuming that no massless  matter fields are
present, since they may certainly introduce (branch-cut) non-analyticities
at zero momentum.

Gauge-invariance plus parity-conservation imply  that the dependence of  the
Lagrangian  on  $A_\mu$  can  only  occur  either through  the   combination
$\fmn=
\partial_\mu A_\nu - \partial_\nu A_\mu$,
or its dual $\tilde F_{\mu}={1\over 2}\epsilon_{\mu\nu\rho}F_{\nu\rho}$.
Each term in the  Lagrangian
must be  built from  contractions of  products of  $\fm$, with  a number of
insertions of the derivative operator $ \partial_\mu $.
For example, if we allow up to four derivatives of the gauge field
$A_{\mu}$, the
Lagrangian can only be a linear combination
of the three following terms
\beq
\fmn \fmn=2\fm\fm, \;\; \fmn \partial^2 \fmn=2 \fm \partial^2 \fm, \;\;
(\fmn \fmn)^2
=4(\fm\fm)^2 \;,
\label{ndptrms}
\eeq
since other contractions vanish due to de Bianchi identity $\partial_{\mu}\fm
=0$.

We will consider initially a more general gauge-invariant action of the form
\bea
S_{inv} &=& \int \, d^3 v \; {\cal L}_{inv} \nonumber\\
{\cal L}_{inv} &=& \frac{1}{4} \fmn \fmn \,+\, \frac{1}{4} \frac{a_1}{M^2}
\fmn g(-\partial^2) \fmn \,+\, \frac{1}{4} \frac{a_2}{M^3} \frac{1}{4!}
(\fmn \fmn)^2 \nonumber\\
&=& \oh {\tilde F}_\mu ( 1 \, + \, a_1 \frac{ g(-\partial^2)}{M^2} )
{\tilde F}_\mu
\,+\, \oh \frac{a_2}{M^3} \frac{1}{4!} ({\tilde F}_\mu {\tilde F}_\mu)^2
\label{sinv}
\eea
where  $a_1$ and $a_2$  are dimensionless functions  of the
(dimensionless)  ratio  $x =  \frac{{\tilde  F}^2}{M^3}= \frac{\fmn \fmn}{2
M^3}$,  $M$  has  the  dimensions  of  a  mass  (note  that  $A_\mu$ has the
dimensions of $M^{\frac{1}{2}}$), and  the numerical factors are  chosen for
convenience.  The constant $M$ along with the functions $a_1, a_2$ and
$g$  are  inputs that
completely  define the
model.

While this is not the most general gauge-invariant
action~\footnote{ For example, terms of the form $(\fm (-\partial^2)\fm)^n,
n>1$,
and $(\fm\partial_{\mu}\fr)^2$ are not included
in Eq. (\ref{sinv})}, it encompasses several important  cases.
Indeed, we wish to interpret the invariant effective action
as coming from integrating  out heavy matter fields. Usually, this integration
cannot be done exactly, and one is forced to use some
approximation. A typical approximation is the derivative expansion
\cite{deriv}, and as
mentioned before, Eq. (\ref{sinv}) reproduces the expansion up to four
derivatives just taking $g(-\partial^2)=-\partial^2$ and constant values
for the functions $a_1$ and  $a_2$. The constant $M$ is naturally
related to the
mass of the heavy
particles, the precise relation and the values of $a_1$ and $a_2$
being of course model-dependent.~\footnote{Note that if  many fields of  the
same kind but with different masses are present, $M$ should be of the  order
of the lighter mass.}

Other approximations can be obtained using heat kernel techniques \cite{heat}:
for strong, slowly varying $\fmn$, the result of the integration of
matter fields gives $a_1=0$ and a model-dependent function $a_2(x)$
\cite{redlich}. In the opposite limit, for weak and rapidly varying
background fields, the action is again given by Eq. (\ref{sinv})
with $a_1=const\,\, , a_2=0$ and \cite{avra}
\beq
g(-\partial^2)\propto \int_0^1 d\xi\,\xi^2 (1+{1-\xi^2\over 4}{(-\partial^2)
\over M^2})^{-1/2}\simeq {1\over 3}-{1\over 60}{(-\partial^2)\over M^2}+...
\label{formfact}
\eeq
Note the non-analyticity of the form factor $g$
in the massless limit.

We now restrict the discussion to the Lagrangian of (\ref{sinv}) with
$g(-\partial^2)=-\partial^2$. In section 4 we will discuss returning to
(\ref{sinv}) with $g$ given by (\ref{formfact}).

To do perturbative calculations in the model defined by (\ref{sinv})
we need to fix the gauge. A particularly convenient choice is to
add to ${\cal L}_{inv}$ in (\ref{sinv}) the gauge-fixing term
${\cal L}_{gf}$:
\bea
{\cal L} &=& {\cal L}_{inv} \;+\; {\cal L}_{gf} \\
{\cal L}_{gf} &=&  \displaystyle{\frac{1}{2 \alpha}}\; \partial_\mu A_\mu ( 1 -
a_1 \frac{\partial^2}{M^2} ) \partial_\nu A_\nu \;,
\label{lgf}
\eea
where $\alpha$ is a gauge-fixing parameter. In particular,
note that for $\alpha = 1$ and $a_1=const$ the gauge field
propagator assumes a `Feynman-gauge'- like form
\beq
\langle A_\mu A_\nu \rangle \;=\; \displaystyle{
\frac{\delta_{\mu \nu}}{k^2
(1 + \frac{a_1}{M^2} k^2)}} \;,
\label{prop}
\eeq
but we shall keep $\alpha$ arbitrary in order to show the gauge-independence
of our results.

Nothing can be concluded about the dimensionless parameters $a_1$ and $a_2$
from symmetry considerations only. However, there are some extra
conditions on the signs of these coefficients. These signs are important in
determining the perturbative properties of the model. A wrong inference
might be drawn by requiring positivity of the Euclidean action $S_{inv}$ for
all the possible field configurations:  that both $a_1$ and $a_2$ must be
positive. But this is not the case, for $a_1$ at least, since a positive
$a_1$ corresponds to {\em anti}-screening of electric charge by vacuum
polarization.  A simple calculation shows that the propagator (\ref{prop})
in coordinate space reads:
\beq
G_{\mu \nu} (r) =  \frac{\delta_{\mu\nu}}{2 \pi r}
( 1 \,-\, e^{-\frac{M r}{a_1}} )
\label{qeff}
\eeq
As the gauge field propagator in the absence of
matter fields is just $\frac{\delta_{\mu\nu}}{2 \pi r}$, the effect of $a_1$
in (\ref{qeff}) is to dress the bare charge $Q$ to the effective one
\beq
Q_{eff} (r) \;=\; Q \; \sqrt{1 \;-\; e^{-\frac{M r}{a_1}}} \;.
\label{qeff1}
\eeq

Thus when $r \to 0$, the effective charge tends to $0$,
corresponding to anti-screening. The conclusion that the correct
choice is a negative $a_1$ was also verified by  explicitly calculating the
coefficient $a_1$ in models containing either bosonic or fermionic fields.

The relevance of the sign of $a_1$ to the properties of the model
is evident from (\ref{prop}); a negative $a_1$ (the realistic
case) produces a pole in the {\em Euclidean \/} gauge field propagator,
located at $k^2 = - \frac{M^2}{a_1}$. Poles in the Euclidean propagators
don't make sense physically (namely, they correspond to particles
with imaginary mass in Minkowski space). This undesirable feature
is excluded in this case by the simple reason that we are dealing with
an effective low-momentum theory,  which we expect to make sense only
up to a certain cutoff momentum . This cutoff is determined by $M$,
and in our case it is enough to take that cutoff smaller than
$\frac{M}{(-a_1)^{1/2}}$ to avoid the unphysical pole.
This means in particular
that loop momenta will run up to a cutoff of that order.

Knowledge of the sign of $a_2$, is not essential to the calculation, and
we might keep it arbitrary. We will, however, assume it to be positive in
order to have a stable theory to start with.

\section{The effective potential}

\subsection{General $a_1$ and $a_2$}

We are concerned with the calculation of the 1-loop effective action in
the case of a constant $\fmn$, with the aim of particularising to the
constant magnetic field situation and constructing an
`effective potential' $V_{eff} (F)$.

The one-loop effective action for the model is easily seen to be given by
\beq
\Gamma_{eff} \;=\; S \;+\; \Gamma^{(1)}
\label{geff}
\eeq
where $S = \int d^3 v {\cal L}_{inv}$, with ${\cal L}_{inv}$ as defined by
(2), and
\beq
\Gamma^{(1)} \;=\; \oh \left. {\rm Tr} \log (\frac{\delta^2 S}{\delta A_\mu
\delta A_\nu}) \,-\, \oh
{\rm Tr} \log (\frac{\delta^2 S}{\delta A_\mu \delta A_\nu})\right|_{A = 0}
\;.
\label{g1}
\eeq
The second functional derivative of $S$, evaluated for a constant $\fmn$
yields
$$ \frac{\delta^2 S}{\delta A_\mu (v) \delta A_\nu (w)} |_{\fmn = const.}
= $$
$$\left\{ \; [ 1 - a_1 \frac{\partial^2}{M^2} +
\frac{2 a'_2}{4!} x^2 + \frac{a_2}{3!} x] \;
(-\partial^2 \delta_{\mu \nu}
+\partial_\mu \partial_\nu)
 - \frac{1 - a_1 \frac{\partial^2}{M^2}
}{\alpha} \partial_\mu \partial_\nu  \right.$$
\beq
\left. -\frac{1}{M^3} \; [ - 4 a'_1 \frac{\partial^2}{M^2} + \frac{a''_2}{3!}
x^2 + \frac{4 a'_2}{3!}  x + \frac{2 a_2}{3!} ]  \;
F_{\mu \alpha} \; F_{\nu \beta}
\partial_\alpha \partial_\beta \right\}
\delta
(v - w) \;,
\label{sfd}
\eeq
where $a'_\alpha = \frac{d a_\alpha}{d x}$, and $v ,\, w$ represent the
coordinates of two spacetime points.

Thus the effective potential for a constant $\fmn$ becomes
\bea
V_{eff} (F) &=& (Vol)^{-1}
[ \;\; S \;+\; \Gamma^{(1)} \;\;]|_{F = constant} \nonumber\\
&=&
V_{cl} (F) \,+\,V^{(1)} (F)  \;.
\label{veffdef}
\eea
Here $V_{cl}$ denotes the part  of the total effective potential that does
not include quantum effects coming from $A_\mu$ (it contains quantum effects
of the matter fields, though)
\bea
V_{cl} (F) &=& (Vol)^{-1}[ S ]_{F = constant} \nonumber\\
&=& [ \;\; \frac{1}{4} \fmn \fmn \;+\; \frac{a_2}{4 M^3} \,
\frac{1}{4!} (\fmn \fmn)^2 \;\; ] \;,
\label{vcl}
\eea
and
$$ V^{(1)} (F) \;=\;(Vol)^{-1} [\;\;\Gamma^{(1)} (F) \;\;]_{F = constant} $$
$$= \oh \, {\rm Tr} \, \log
\left\{
[1- a_1 \frac{\partial^2}{M^2} \,+\,\frac{2 a'_2}{4!}
x^2\,+\, \frac{a_2}{3!}x](-\partial^2 \delta_{\mu\nu}
+\partial_\mu \partial_\nu)\right.$$ $$ \left.
-\frac{1}{\alpha}\partial_\mu
\partial_\nu -\frac{1}{M^3} [-4 a'_1 \frac{\partial^2}{M^2} +
\frac{a''_2}{3!} x^2 + \frac{4 a'_2}{3!} x + \frac{2
a_2}{3!}]F_{\mu\alpha} F_{\nu\beta} \partial_\alpha \partial_\beta
\right\}$$
\beq
-\oh {\rm Tr} \log \left\{
(1-\frac{a_1(0)}{M^2} \partial^2)(-\partial^2 \delta_{\mu\nu} +\partial_\mu
\partial_\nu) - \frac{1- \frac{\partial^2}{M^2}}{\alpha}
\partial_\mu \partial_\nu
\right\} \;.
\label{v1}
\eeq

In what follows we shall evaluate $V^{(1)}$, the one-loop correction to the
classical potential. It is convenient to convert first the functional trace
in (\ref{v1}) to momentum space, and then to introduce the field-dependent
projectors defined in the appendix. This procedure yields
\beq
V^{(1)}(F) \;=\; \oh \, \int_\Lambda \frac{d^3 k}{(2 \pi)^3}
\; {\rm tr} \, \log [ P_{\mu\nu} \,+\, \alpha_Q \; Q_{\mu \nu} \,+\, \alpha_R
\; R_{\mu\nu} ] \;,
\label{v1ms}
\eeq
where
$$\alpha_Q \;=\; \frac{1+a_1
\displaystyle{\frac{k^2}{M^2}} +
\displaystyle{\frac{a_2}{3!}} x
+
\displaystyle{\frac{2 a'_2}{4!}} x^2}{1 + a_1 (0)
\displaystyle{\frac{k^2}{M^2}}} $$
\beq
\alpha_R \;=\; \alpha_Q \,+\, \frac{
\displaystyle{\frac{2 a_2}{3!}} + 4 a'_1
\displaystyle{\frac{k^2}{M^2}} +
\displaystyle{\frac{4 a'_2}{3!}} x
+
\displaystyle{\frac{a''_2}{3!}} x^2}{M^3(1 + a_1 (0)
\displaystyle{\frac{k^2}{M^2}})}
({\tilde F}^2 - \frac{(k \cdot {\tilde F})^2}{k^2}) \;.
\eeq
The subscript $\Lambda$ in the momentum integration means that a Euclidean
cutoff $\Lambda \sim M$ has been introduced. This cutoff $\Lambda$ takes
care of the fact that high momentum fluctuations of the gauge field are not
described by the classical effective theory, so they should not be included in
the loops for the quantum effective theory derived therefrom. A precise
value for $\Lambda$ cannot be decided {\em a priori \/}. It certainly should
be of the order of $M$, and one expects the final results to be insensitive
to small changes in $\Lambda$ around $M$.

The evaluation in (\ref{v1ms}) of the trace over Lorentz indices
of an involved function of an $\fmn$-dependent tensor is made possible
by the use of the complete set of orthogonal projectors introduced
in the Appendix, where we explain this
construction and its application to the present case. The
result of taking the trace over Lorentz indices may be written as
$$ V^{(1)}(F) \;=\;
V^{(1)}_a (F)\;+\; V^{(1)}_b (F) $$
$$ V^{(1)}_a (F) \;=\; \oh
\int_\Lambda \frac{d^3 k}{(2 \pi)^3} \, \log [ 1 \,+\,
\frac{ (a_1-a_1(0)) \displaystyle{\frac{k^2}{M^2}} +
\displaystyle{\frac{a_2}{3!}}
x + \displaystyle{\frac{2 a'_2}{4!}} x^2}{
1 + a_1(0) \displaystyle{\frac{k^2}{M^2}} } ] $$

$$V^{(1)}_b (F) \;=\; \oh  \int_\Lambda \frac{d^3 k}{(2 \pi)^3} \, \log [ 1
\,+\,
\frac{(a_1 - a_1(0)) \frac{k^2}{M^2} + \frac{a_2}{3!} x+
\frac{2 a'_2}{4!} x^2}{1 + a_1(0) \displaystyle{\frac{k^2}{M^2}}}+$$
\beq
+
\frac{({\tilde F}^2 - \frac{(k \cdot {\tilde F})^2}{k^2})}{M^3(1+
 a_1 (0)\frac{k^2}{M^2})} \left(
(a_1 - a_1 (0)) \frac{k^2}{M^2} +\frac{2 a_2}{3!} + 4 a'_1 \frac{k^2}{M^2}
+ \frac{4 a'_2}{3!}x +  \frac{a''_2}{3!} x^2
\right)
] \;.
\label{loop}
\eeq

As $a_1$ is negative, we redefine:
$a_1 \,\to \, - a_1$, and shall consider $a_1$ positive in what follows.
Before dealing with the evaluation of the integrals in (\ref{loop}), it is
worth studying an important aspect of the one loop correction (\ref{loop}),
namely, whether it can be negative. A negative $V^{(1)}$ is a necessary
condition for the existence of instabilities, which may shift the true vacuum
to a non-zero value of $\fmn$. It is not difficult to check that, if
$a_1$ and $a_2$ are constants, the correction $V^{(1)}$ is always
positive, since its sign is determined by integrals of logs of functions
larger than $1$ (of course we are assuming  $a_2 \geq 0$). In the general
case of field-dependent coefficients, requiring positivity of $V^{(1)}$
yields conditions on the coefficients and their derivatives. Sufficient
(but not necesary) conditions to ensure stability are, for example 
\bea
a_1 (x) & \leq & \; a_1 (0) \nonumber\\
a_2 (x) & = & \; C \; x^{\alpha} \;\;\; , \;\; C > 0 \;\; , \;\; \alpha > - 2 \;.
\eea
The above conditions are quite likely to be met in most of the situations.
The first
one ensures that the spurious pole does not move towards the origin when
the field
grows, and the second one says that the `interaction term'
(proportional to $a_2 \, x^2$)
in the effective Lagrangian does not tend to zero for large $x$.

It remains then to evaluate the cutoff
momentum integrals in (\ref{loop}).
These integrals cannot be exactly evaluated as they stand. However
some more explicit results may be provided under different
simplifying assumptions,
which we describe next.

\subsection{The case of constant $a_1$ and $a_2$}

This situation corresponds to assuming that $a_1$ and $a_2$ bear
no dependence on the field $\fmn$, and thus become just constant
parameters. This is the case when the classical effective action
is expanded up to four derivatives of $A_{\mu}$.
Defining the two dimensionless parameters $\gamma$ and $y$,
\beq
\gamma \;\;=\;\; a_1^{\frac{1}{2}} \frac{\Lambda}{M} \;\;,\;\; y \;=\;
\frac{a_2 {\tilde F}^2}{3 ! M^3} \;=\; \frac{a_2}{3!} x \;.
\label{para}
\eeq
the results of the integrations may be put as
$$ V^{(1)}_a \;=\; \frac{M^3}{3 (2 \pi)^2 a_1^{\frac{3}{2}}} \; \left\{
\gamma^3 \log ( 1 + y ) \,-\,  2 \gamma y \;+\; \gamma^3 \log [ 1
\,-\,\frac{\gamma^2}{1 + y}] \right.$$
\beq
\left. + (1 + y)^{\frac{3}{2}} \, \log [\frac{(1+y)^{\frac{1}{2}}
\,+\,\gamma}{(1+y)^{\frac{1}{2}} \,-\,\gamma} ]\,- \, \gamma^3 \, \log ( 1 -
\gamma^2 ) \,-\, \log( \frac{1+\gamma}{1-\gamma} ) \,\right\}
\label{v1a}
\eeq
and
\bea
V^{(1)}_b &=& \frac{M^3}{3 (2 \pi)^2 a_1^{\frac{3}{2}}} \; \left\{ -
\gamma^3  \, \log ( 1 - \gamma^2 ) - \, \log(\frac{1+\gamma}{1-\gamma})
\right. \nonumber\\ &+& \gamma^3 \log (1 + y)
\,-\, 2 \gamma^3 \,+\, \gamma^3 \log ( 1 - \frac{\gamma^2}{1+y} )
\nonumber\\ &+& \gamma^3 \sqrt{\frac{2 (1- \gamma^2 + 3 y)}{y}} \,
{\rm arctanh} \, \sqrt{\frac{2 y}{(1-\gamma^2 + 3 y)}} \nonumber\\
&-& \left. \frac{14}{3} \gamma \, y  \,+\, \int_0^1 d s [ 1 + y (3 - 2 s^2)
]^{\frac{3}{2}} \;  \log [ \frac{(1 +  y
(3 - 2 s^2) )^{\frac{1}{2}} \,+\, \gamma}{(1 +  y (3 - 2 s^2)
)^{\frac{1}{2}} \,-\, \gamma }] \right\} \;\;.
\label{v1b}
\eea
This completes the computation of $V^{(1)}$, where all but the last term in
$V^{(1)}_b$ of (\ref{v1b}) are given in a closed analytic form.

\subsection{The case $a_1 = 0$}

If the quantum matter fields are  charged fermions of mass $M$,
the invariant Lagrangian
(resulting from integrating out the fermions) can be computed
exactly for a {\it constant} $\fmn$.
It is given by \cite{redlich}
\bea
{\cal L}_{inv}&=&{1\over 4}\fmn\fmn-
\frac {M^3 u^{3/2}}{8\pi^{3/2}}\int_0^{\infty}
d\xi \xi^{-5/2}[\xi coth[\xi] - 1]e^{-\frac{\xi}{u}}\nonumber\\
&\equiv & {1\over 4}\fmn\fmn+\Delta{\cal L}
\label{lexac}
\eea
where $u=\frac{eB}{M^2}=e\left (\frac{2x}{M}\right )^{1/2}$. The integral in
the above expression can be readily computed
and gives
\beq
\Delta{\cal L}=\frac{M^3}{8\pi}[\frac{4}{3}-2u+4 (2u^3)^{1/2}\zeta(-1/2,1/2u)]
\label{riemann}
\eeq
where $\zeta$ denotes the generalized Riemann zeta-function.

When $u\ll 1$, the above expression reproduces the
Schwinger DeWitt (or inverse mass) expansion of the effective action
for a constant background.
In the opposite limit, when $u\gg 1$, $\Delta {\cal L}$ is proportional
to $(eB)^{3/2}$. In any case, the exact Lagrangian is of the form
Eq. (\ref{sinv}) with $a_1=0$. The function $a_2(x)$ can be easily
obtained by comparing Eqs. (\ref{sinv}) and (\ref{lexac}).

In this situation,
the results of the integrals for $V^{(1)}_a$ and $V^{(1)}_b$ are
combined to yield
\beq
V^{(1)} \;=\; \frac{\Lambda^3}{6  \pi^2} \left\{ \log [ 1 +
\frac{x}{3!} (\frac{a'_2}{2} x + a_2)] + \rho \, {\rm arctanh}
\rho^{-1} -1 \right\} \;,
\eeq
where
\beq
\rho = \sqrt{ \frac{1+
\displaystyle{\frac{x}{3!}} (3 a_2 + \frac{9 a'_2}{2} x
+a''_2 x^2)}{\frac{x}{3!} ( 2 a_2 + 4 a'_2 x + a''_2 x^2)}}
\eeq

\section{Discussion}
The use of a cutoff $\Lambda$ in the momentum integrals
corresponding to gauge-field
propagators was necesary to avoid reaching the unphysical pole at
Euclidean momentum.
This pole appears because  the classical effective action has
been truncated to contain no more than two derivatives
acting on $\fmn$, leading to
a polynomial with two zeroes in $k^2$, hence a propagator with two poles.
On the other
hand one may verify that, if no momentum truncation is made, the resulting
propagator has only one
pole, at the physical point $k^2 = 0$ (see Eq.(\ref{formfact})).
Thus a possible way out of the problem of the presence of the
spurious pole would be
to keep the full momentum dependence of the quadratic
part~\footnote{Of course, this
procedure may be criticized on the grounds that it treats differently the
quadratic part and the interaction terms.}, which then becomes
non-local in coordinate space. More explicitly, instead of using
the truncation $g(-\partial^2)=-\partial^2$ one can work with
the complete form factor defined in Eq. (\ref{formfact}).
In this case,
the one loop correction to the potential becomes
$$V^{(1)} \;=\; V^{(1)}_a \,+\, V^{(1)}_b$$
$$V^{(1)}_a \;=\; \frac{M^3}{(2 \pi)^2} \, \int_0^{\frac{\Lambda}{M}} dq q^2
\log \left[ 1 \,+\, \frac{ \displaystyle{\frac{1}{3!}} ( \displaystyle{
\frac{a'_2}{2}} x^2 + a_2 x) }{1 + a_1 g(q^2)} \right] $$
\beq
V^{(1)}_b \;=\; V^{(1)}_a \;+\; \frac{2 M^3}{(2 \pi)^2}
\int_0^{\frac{\Lambda}{M}}
dq \; q^2 \left( \rho \; {\rm arctanh} \rho^{-1} \;-\;1 \right)
\eeq
where now
\beq
\rho \;=\; \sqrt{ \frac{1+a_1 g+4 a'_1 g+ \frac{1}{3!} (a''_2 x^3 + \frac{9}{2}
a'_2 x^2 + 3 a_2 x)}{x [ 4 a'_1 g + \frac{1}{3!} (a''_2 x^2 + 4 a'_2 x
+ 2 a_2)]}}\;.
\eeq
and it is possible to take the limit $\Lambda\rightarrow\infty$.

It is important to remember that even if the initial theory is renormalizable,
the effective theory is not necesarily so, as in the case of the Fermi
theory regarded as (part of) an effective theory for the
renormalizable Electroweak
Lagrangian. However, had we known the initial renormalizable theory, we
would have a recipe to absorb the possible infinities, by means of
counterterms which in some
cases are non-local or non-polynomial in terms of the effective theory
variables.

We would now like to check whether our results for the one-loop
correction to the effective potential are very sensitive to the precise
value of the cutoff $\Lambda$ or not. We would of course like to have at
least some regime where the actual value of $\Lambda$ is not so relevant.
This might be done by studying the derivative of $V^{(1)}$ with respect to
$\Lambda$, to see if it is small. But as both $V^{(1)}$ and
$\Lambda$ are dimensionful quantities, that smallness  should be considered as
relative to another dimensionful quantity, the natural one being the mass $M$.
This amounts to considering the magnitude of
$\displaystyle{\frac{\partial  v^{(1)} }{\partial \frac{\Lambda}{M}}}$
where we have introduced the dimensionless potential $v^{(1)} =
\frac{V^{(1)}}{M^3}$. It is easy to realize that in our results
$V^{(1)} \propto \Lambda^3$ which implies:
$\displaystyle{\frac{\partial  v^{(1)} }{\partial \frac{\Lambda}{M}}}
\propto (\frac{\Lambda}{M})^2$. The outcome of this simple estimate is
that the results are not very sensitive to the actual value of $\Lambda$
when $\displaystyle{\frac{\Lambda}{M}} << 1$, since the derivative is
small. If we want to integrate momenta which are nearer to $M$, more
higher order terms in the derivative expansion should be included.

It is worth noting that the gauge-fixing  independence of the results
is explicit, since $\alpha$ does not appear in the expressions for
$V^{(1)}$. This is a consequence of the fact that the effective
action is gauge-fixing independent on-shell, and certainly a constant
$\fmn$ satisfies the  classical equations of motion.

We finally remark that the  assumption of parity conservation was essential in
determining the form of the classical effective theory,
and we speculate that the results of performing an
analogous calculation in the parity  violating case may
be qualitatively different, because in the latter
case there  will be a Chern-Simons term inducing a
topological mass for the gauge field, which will strongly
modify the low-momentum regime of the gauge-field
propagator. To the mass $M$ one should then add the new scale
set by the topological mass.

\section {Acknowledgments}
C.~D.~F. acknowledges the Department of Physics of the Univ. of
Oxford for its kind hospitality.
F.~D.~M. enjoyed the hospitality of the General Relativity Group at the
University of Maryland, where part of this work was done.
The work of F.~D.~M. was supported by CONICET, UBA and Fundaci\' on
Antorchas.

\newpage

\appendix
\section{}

We construct here the three orthogonal projectors used in the calculation
of the effective potential. We have written $V^{(1)}$ in Equation
(\ref{v1ms}) as
\beq
V^{(1)} (F) \;=\; \oh \int_\Lambda \frac{d^3 k}{(2 \pi)^3} \,
{\rm tr} \, \log \, \left[\alpha_P (k^2) P_{\mu \nu} \,+\,\alpha_Q (k^2)
Q_{\mu \nu} \,+\, \alpha_R (k^2) R_{\mu \nu} \right] \;,
\label{a1}
\eeq
where $P$, $Q$ and $R$ verify
$$
P^2 \; = \; P \;\;,\;\; Q^2 \;=\; Q \;\;,\;\; R^2 \;=\; R
$$
\beq
P \,+\,Q\,+\,R\;=\; 1
\label{a2}
\eeq
and all the products between different projectors vanish.
Using the properties (\ref{a2}), we can write (\ref{a1}) as
$$ V^{(1)} (F) \;=\; \oh \int_\Lambda \frac{d^3 k}{(2 \pi)^3} \,
{\rm tr} \, \log \, \left[ \alpha_P (k^2) P_{\mu \nu}
\,+\,\alpha_Q (k^2)  Q_{\mu \nu} \,+
\, \alpha_R (k^2) R_{\mu \nu} \right]$$
$$
= \;\oh {\rm tr} \,P\,
\int_\Lambda \frac{d^3 k}{(2 \pi)^3}\,\log \,( \alpha_P (k^2) )
\,+\, \oh {\rm tr} \,Q\,
\int_\Lambda \frac{d^3 k}{(2 \pi)^3}\,\log \,( \alpha_Q (k^2) )
$$
\beq
+\, \oh {\rm tr} \,R\,
\int_\Lambda \frac{d^3 k}{(2 \pi)^3}\,\log \,(\alpha_R (k^2) )
\;.
\label{a3}
\eeq
Taking into account that the only possible eigenvalues for $P$, $Q$ and $R$
are $0$ and $1$, and that they are $3 \times 3$ matrices, we may also
deduce from (\ref{a2}) that
\beq
{\rm tr} \, P \;=\;{\rm tr}\,Q\;=\; {\rm tr}\,R \;=\;1 \;.
\label{a4}
\eeq

To find the functions $\alpha_P$, $\alpha_Q$
and $\alpha_R$, we chose the set of projectors:
\bea
P_{\mu\nu} &=& \displaystyle{\frac{k_\mu k_\nu}{k^2}} \nonumber\\
Q_{\mu\nu} &=& \frac{k^2 {\tilde F}_\mu {\tilde F}_\nu +
(k \cdot {\tilde F})^2 \displaystyle{\frac{k_\mu k_\nu}{k^2}} -
k \cdot {\tilde F} (k_\mu {\tilde F}_\nu + k_\nu {\tilde F}_\mu)}{
k^2 {\tilde F}^2 - (k \cdot {\tilde F})^2} \nonumber\\
R_{\mu\nu} &=& \frac{(k^2 {\tilde F}^2 - (k \cdot {\tilde F})^2)
\delta_{\mu\nu} - {\tilde F}^2 k_\mu k_\nu - k^2 {\tilde F}_\mu
{\tilde F}_\nu
+ k \cdot {\tilde F} (k_\mu {\tilde F}_\nu + k_\nu {\tilde F}_\mu)}{
k^2 {\tilde F}^2 - (k \cdot {\tilde F})^2}  \;.
\label{a5}
\eea

\end{document}